\documentclass[english, journal=jcisd8, manuscript=article,
  layout=twocolumn]{achemso} \SectionNumbersOn
\usepackage[T1]{fontenc}
\usepackage[latin9]{inputenc}
\usepackage{verbatim}
\usepackage{cprotect}
\usepackage{booktabs}
\usepackage{amstext}
\usepackage{graphicx}
\usepackage{siunitx}
\usepackage{xr}
\DeclareSIUnit\angstrom{\text{\AA}} \DeclareSIUnit[number-unit-product =
{\,}]\cal{cal} \DeclareSIUnit\kcal{\kilo\cal}

\makeatletter

\author{Felipe Silva Carvalho}

\affiliation{Department of Physics and Astronomy, California State University,
Northridge, Northridge, CA 91330}

\author{Steven Ramsey}

\affiliation{Department of Chemistry, Lehman College, Bronx, NY 10468}

\author{George M. Giamba\c{s}u}

\affiliation{Department of Chemistry and Chemical Biology, Rutgers, The State University of New Jersey, New Brunswick, New Jersey 08901, United States}
\altaffiliation{Discovery Chemistry, Modeling \& Informatics, Merck Research Laboratories, Merck \& Co., Inc., 33 Avenue Louis Pasteur, Boston, MA 02115}

\author{Tom Kurtzman}

\affiliation{Department of Chemistry, Lehman College, Bronx, NY 10468}
\email{simpleliquid@gmail.com}

\author{Tyler Luchko}

\affiliation{Department of Physics and Astronomy, California State University,
Northridge, Northridge, CA 91330}

\email{tluchko@csun.edu}

\title{Solv-eze: Automated Placement of Explicit Water Molecules Using 3D-RISM}

\abbreviations{}


\providecommand{\tabularnewline}{\\}

\providecommand*{\code}[1]{\texttt{#1}}

\usepackage{threeparttable}
\usepackage{pdflscape}
\usepackage[bottom]{footmisc}

\makeatother

\setkeys{acs}{keywords=true}

\usepackage{babel}
\begin{document}

\begin{abstract}
Molecular dynamics (MD) simulations are widely used to study biological systems,
where water molecules often play a critical role in protein-ligand interactions.
In conventional MD preparation protocols, water molecules are typically added
from a pre-equilibrated solvent box and removed using conservative steric
cutoffs, an approach that can eliminate important interfacial waters that are
often not recovered during equilibration due to kinetic barriers limiting
exchange with bulk solvent. In this work, we present an automated and
computationally efficient method for placing water molecules around biomolecular
solutes using three-dimensional reference interaction site model (3D-RISM)
solvent density distributions. By identifying regions of high solvent
probability, the method generates physically meaningful initial hydration
structures without requiring extended sampling or specialized techniques such as
grand canonical Monte Carlo (MC) or hybrid MC/MD approaches, and will be
released as an update to AmberTools 26, enabling seamless integration into
standard MD preparation pipelines. We validated the approach on a diverse set of
protein-ligand complexes with crystallographically resolved bridging waters,
showing that the method reproduced over 80\% of experimentally observed bridging
waters and 85\% of buried waters not accessible to the bulk. Subsequent
energy minimization of both crystallographic and predicted waters further improved 
agreement. Overall, this
method enables more accurate and practical initialization of interfacial
hydration, improving the reliability of MD simulations with modest computational
cost relative to routine system preparation.
\end{abstract}

\section{Introduction}

Molecular dynamics (MD) simulation is a fundamental tool for studying
biomolecular systems, providing atomic-level insights into structure, dynamics,
and
thermodynamics.\cite{filipe2022molecular,bittner2024investigating,wu2022application}
The validity of thermodynamic properties computed from MD trajectories relies on
the ergodic hypothesis\cite{birkhoff_proof_1931}, which assumes that time
averages over sufficiently long trajectories converge to ensemble averages.
However, practical MD simulations are necessarily finite, and many biologically
relevant processes occur on timescales far exceeding accessible simulation
lengths. This fundamental limitation poses significant challenges for systems in
which slow relaxation processes govern the equilibrium distribution of molecular
configurations.

Water molecules play an essential role in virtually all aspects of biomolecular
function.\cite{wiggins1990role,dargaville2022water,ball2017water,bellissent2016water}
In protein-ligand binding, water molecules at the interface can mediate
recognition through hydrogen-bonding networks, contributing significantly to
binding affinity and specificity.\cite{zsido2025water} Crystallographic studies
frequently reveal ordered water molecules that form bridging hydrogen bonds
between protein and ligand atoms, underscoring their structural importance in
molecular recognition.\cite{Levy2006,darby2019water} 

Standard protocols
\cite{roe_protocol_2020,abraham_gromacs_2015,joCHARMMGUIWebbasedGraphical2008,braun_best_2019,case2023ambertools}
for preparing solvated biomolecular systems involve overlaying the solute with a
pre-equilibrated water box and subsequently deleting water molecules that
sterically overlap with solute atoms. The overlap criteria typically uses van der Waals 
radii, thereby overestimating the region inaccessible to water. 
Furthermore, the discrete water molecules in the pre-equilibrated box are unlikely 
to coincide with favorable hydration sites, particularly in buried cavities and 
protein-ligand interfaces. This conservative approach
artificially creates regions of zero water density, or vacuum, on the solute
surface, at protein-ligand interfaces, and in buried cavities. Bridging water
sites at protein-ligand interfaces are of particular concern, as the deletion
protocol systematically removes water molecules that should mediate protein-ligand
interactions.

After preparatory steps, the
system undergoes energy minimization and is relaxed through MD simulation. These
steps are employed with the expectation that water molecules will diffuse into
underpopulated regions and establish distributions consistent with the target
ensemble. However, this assumption breaks down for water molecules in occluded
environments such as structural waters buried within proteins, waters in gated
cavities requiring conformational changes for exchange, and, critically,
bridging waters between proteins and their bound
ligands.\cite{wall_biomolecular_2019,williams1994buried,Levy2006,darby2019water,ben-shalom_simulating_2019}
For buried bridging water sites to properly equilibrate through conventional MD,
the ligand would need to unbind to permit water exchange, a dynamic process with
timescales often orders of magnitude longer than practical classical MD
equilibration simulations.\cite{pal2004dynamics}  

Many computational approaches have been developed to address the challenge of
water equilibration in buried or occluded sites (Zsid\'{o} and Het\'{e}nyi
provide an extensive list \cite{zsido2025water}). Grand canonical Monte
Carlo\cite{widom_topics_1963,mezei_cavity-biased_1980,perego_molecular_2015}
(GCMC) is a physics-based method that enables water insertion and deletion moves
to sample variable water occupancy
states\cite{ge_enhancing_2022,deng_efficient_2024,samways_grand_2020}, which can
be integrated with molecular dynamics simulations. Ben-Shalom et al.
\cite{ben-shalom_simulating_2019} introduced a hybrid MC/MD approach in
Amber\cite{caseRecentDevelopmentsAmber2025} that combines translational Monte
Carlo water moves with MD to exchange water between bulk and buried sites. While
these methods are thermodynamically rigorous, they require substantial
computational resources, specialized simulation protocols, and careful parameter
selection, making them impractical for rapid or automated simulation setup.
Recently, several machine learning approaches have been
developed,\cite{huangAccuratePredictionHydration2021,parkGalaxyWaterCNNPredictionWater2022,zamanosHydraProtNewDeep2024,kuangSuperwaterGenerativeAI2025,liHydraMapV2Prediction2023}
which are trained on crystal structures that include waters. While these methods
can be very fast, they are fundamentally limited by the quantity and quality of
waters modeled into crystal structures.

The three-dimensional reference interaction site
model\cite{beglov_integral_1997,chandler_optimized_1972,hirata_extended_1981,hirata_three-dimensional_2004,kovalenko_three-dimensional_1998}
(3D-RISM) theory offers an attractive alternative for characterizing the initial
water distributions around biomolecules. As a statistical mechanical theory of
solvation, 3D-RISM computes three-dimensional solvent density distributions that
incorporate both solute-solvent and solvent-solvent interactions using the same
force fields employed in MD simulations. Importantly, 3D-RISM provides
equilibrium solvent distributions without requiring dynamic equilibration,
making it well-suited for identifying high-probability water positions in
regions inaccessible to standard simulation protocols. 

However, 3D-RISM only provides water density distributions, which must be
processed to place water molecules.  Placevent is an iterative method that uses
local maxima in the water density distribution to place waters
\cite{sindhikara_placevent:_2012}. GAsol also identifies hydration sites from
the water density distribution, but then uses a genetic algorithm to optimize
the water network.\cite{fusaniOptimalWaterNetworks2018} Laplacian mapping (LM) is a third method
that uses the curvature of the water density distribution to identify hydration
sites.\cite{case2023ambertools,giambasu2019predicting}

Here we present Solv-eze, an automated tool for placing water molecules around
biomolecular solutes based on 3D-RISM solvent density distributions processed by 
LM, Placevent or GAsol. Using a physics-based approach, our method systematically 
places water
molecules at appropriate positions, providing initial configurations that better
approximate equilibrium solvation. Unlike GCMC or hybrid MC/MD methods, our
approach requires no extended sampling. The generation of water distributions
takes several minutes for the solvation of typical proteins. The method is fully
automated and will be released as an update to AmberTools 26, making it
accessible to researchers preparing biomolecular
simulations\cite{case2023ambertools}. The approach was validated against 93
protein-ligand systems, which contained bridging waters in the crystallographic
structure.\cite{lu2007analysis}

The remainder of this paper is organized as follows. Section 2 provides an
overview of the Solv-eze workflow, water placement methods, and their use within
standard simulation setup. Section 3 describes the validation methodology.
Section 4 presents the results, including the effects of the LM
threshold, energy minimization, and computational cost. We conclude by
summarizing the key advantages of Solv-eze water placement for establishing
physically meaningful initial solvation configurations.

\section{Software overview}

\subsection{Water placement workflow\label{subsec:packaga-Water-placement}}

\begin{figure}
\includegraphics{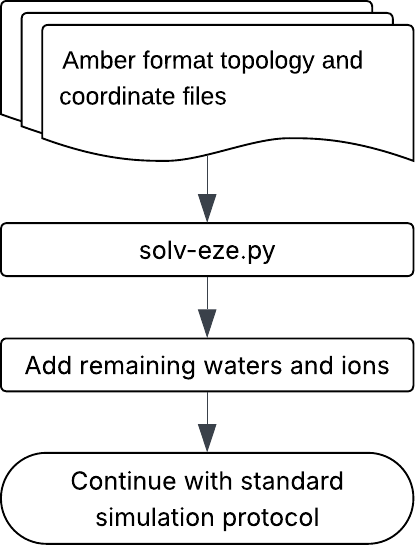}

\caption{User workflow.\label{fig:User-workflow.}}
\end{figure}

Solv-eze automates the placement of explicit water molecules on the surfaces of
biomolecules based on RISM theory as part of a standard system preparation
protocol (Figure \ref{fig:User-workflow.}). First, the user creates parameter
and coordinate files with periodic boundaries in the Amber format. These files
can be constructed with any software package that supports output in the Amber
format, such as \code{tleap}\cite{case2023ambertools},
CHARMM-GUI\cite{joCHARMMGUIWebbasedGraphical2008,leeCHARMMGUIInputGenerator2016},
BioSimSpace\cite{hedgesBioSimSpaceInteroperablePython2019},
CPPTRAJ\cite{roePTRAJCPPTRAJSoftware2013}, or
ParmEd\cite{ParmEdParmEd2025,shirtsLessonsLearnedComparing2017}. Only
coordinates, Lennard-Jones parameters, and atomic partial charges are required.
Solv-eze then creates a PDB file with high-probability water molecules, to which
remaining water and co-solvent molecules can be added, and the system can be
parameterized using standard tools.

\begin{figure*}
\includegraphics{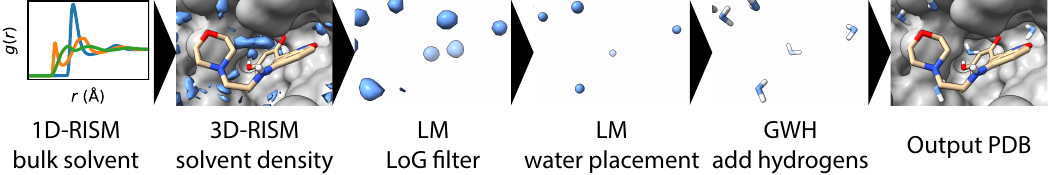}

\caption{Overview of the package workflow. Panel 1: Site-site pair correlation
functions of water: O-O (blue), O-H (orange), and H-H (green). Panel 2: Protein
structure (grey surface) and ligand molecule (licorice representation; hydrogen:
white, oxygen: red, phosphorus: orange, carbon: green, nitrogen: blue) from heat
shock protein 90 (Hsp90)
(PDB ID: 3BMY)\cite{gopalsamyDiscoveryBenzisoxazolesPotent2008a}, along with the
solute-solvent oxygen density distribution (blue isosurface) obtained from
3D-RISM. Panel 3: LoG filter of the oxygen density distribution (blue isosurface).
Panel 4: Predicted water oxygen atom positions (blue). Panel 5: Reconstructed
water molecules (oxygen: blue; hydrogen: white). Panel 6: Final result showing
the protein, ligand, and predicted water molecules, using the same color
scheme as in the previous panels.\label{fig:Package-workflow.}}

\end{figure*}

The Solv-eze workflow, represented as a diagram in Figure
\ref{fig:Package-workflow.}, begins by solving the 1D-RISM equations for pure
water to generate the bulk solvent susceptibility. If the default values for
1D-RISM are used, the package will use a precalculated solvent susceptibility
result stored in a \code{.xvv} file. The solvent susceptibility, parameter, and
coordinate files are then input into
3D-RISM\cite{gray2022integral,case2023ambertools,luchkoThreeDimensionalMolecularTheory2010},
which computes the 3D number-density distribution of the water oxygen and
hydrogen atoms about the solutes.  Once the number-density distributions are
solved, the package employs LM through
\code{metatwist}\cite{case2023ambertools,giambasu2019predicting}, Placevent
through \code{placevent}\cite{sindhikara_placevent:_2012}, or GAsol through
\code{gasol}\cite{fusaniOptimalWaterNetworks2018} to analyze the oxygen site
distribution function and place water oxygen atoms. Finally, the guess water
hydrogen (\code{gwh}) tool is used to generate plausible orientations of the
water molecules by assigning initial hydrogen
positions.\cite{case2023ambertools} The resulting structures can then be used as
starting configurations for molecular dynamics simulations.

\subsection{Water placement methods}

\subsubsection{Placevent}

The Placevent method\cite{sindhikara_placevent:_2012} begins by converting the
water oxygen density distribution to a population distribution,
\begin{equation}
    P(\vec{r}) = V_\text{voxel}\rho(\vec{r})\label{eq:pop_dist_func},
\end{equation}
where $V_\text{voxel}$ is the volume of a voxel, and $\rho(\vec{r})$ is the
water-oxygen number density at location $\vec{r}$. A water oxygen is first
placed at the global maximum of this population distribution. The population
associated with one oxygen atom is then removed from the surrounding voxels,
conserving the total removed population while depleting the local density near
the placed site. This placement-and-depletion procedure is repeated until the
original water oxygen number density at the next selected site is below a
user-defined cutoff, which defaults to 1.5 times the bulk water density used in
the 3D-RISM calculation.

\subsubsection{GAsol}

GAsol\cite{fusaniOptimalWaterNetworks2018} uses the oxygen number density
distribution from 3D-RISM to identify possible hydration sites, and then
optimizes the positions of the water oxygens on this grid. Candidate hydration
sites are those whose oxygen number density exceeds a threshold  (defaults to 5
times the bulk density of water). This set may be further restricted to a sphere
centered on the ligand or a user-defined point. Each site is then assigned a
value, 
\begin{equation}
    g^\text{max}_i = P_i/r_i
\end{equation}
where $P_i$ is the integrated local water population around site $i$, obtained
by integrating the population distribution over increasingly large spheres until
$P_i \approx 1$ or a maximum radius is reached, and $r_i$ is the corresponding
sphere radius.

A population of trial water distributions is then generated and optimized. Each
individual is a binary vector of candidate grid point sites $\vec{s}$, with
$s_i=1$ indicating the presence of a water molecule and $s_i=0$ indicating its
absence. The density contribution for an individual is calculated as
\begin{equation}
    d_1 = \frac{\sum_i s_i g^\text{max}_i}{\sum_i g^\text{max}_i}.
\end{equation}
Pairs of selected sites separated by less than approximately \SI{3}{\angstrom}
are treated as incompatible. The fitness is then calculated as
\begin{equation}
    F = \sqrt{d_1 d_2} - p,
\end{equation}
where $d_2=1$ if no incompatible pairs are selected and $d_2=0$ otherwise, and
$p$ is  the number of incompatible selected pairs divided by the number of
candidate sites. The fitness of the population is maximized using a genetic
algorithm.

\subsubsection{Laplacian mapping}

LM has been included in the MoFT library since version 18 of the AmberTools suite. 
When accessed through the \code{metatwist} command-line interface, it provides tools for
analyzing 3D particle distributions obtained from 3D-RISM, molecular dynamics,
and other molecular simulations, or experimental methods such as cryo-EM/ET and
X-ray crystallography.

LM uses the Laplacian of particle distributions to locate, define, and map particle binding modes. This approach
is analogous to Bader analysis in atoms-in-molecules methods applied to electron
densities from electronic structure calculations,\cite{baderAtomsMolecules1985}
and more broadly to scale-space blob-detection methods used to identify
localized structures in scalar
fields\cite{lindebergFeatureDetectionAutomatic1998}.

First, a Laplacian-of-Gaussian (LoG) filter is applied to the oxygen number
density distribution using the kernel
\begin{equation}
K(\mathbf{r}; \sigma)
=
\exp\left(-\frac{|\mathbf{r}|^2}{2\sigma^2}\right)
\frac{|\mathbf{r}|^2 - 3\sigma^2}
{2\sqrt{2}\,\pi^{3/2}\,\sigma^7},
\end{equation}
where $\sigma$ is the width of the Gaussian used to smooth the density
distribution. For water placement, we use $\sigma=\SI{1}{\angstrom}$.

Regions where the LoG-filtered density is negative correspond to locally
concentrated density features. To focus on more site-bound modes, LM
introduces a positive sub-unitary threshold factor $t$ and retains only grid
points satisfying
\begin{equation}
L(\mathbf{r}) < t L_{\min},
\end{equation}
where $L_{\min}$ is the minimum value of the LoG-filtered map. This selects the
most negative part of the LoG response, i.e., the strongest locally concentrated
regions.

LM then groups contiguous retained voxels into connected components, or blobs,
using periodic connectivity. Finally, water oxygen positions are placed at the
local minima of the LoG-filtered map within each blob. Thus, a blob may generate
more than one placed water if it contains multiple local minima.

\subsection{Solv-eze usage}

\begin{table*}
\caption{List of required and some of the optional arguments necessary for
running the package.\label{tab:List-of-options}}

\begin{tabular}{>{\raggedright}p{1.25in}>{\raggedright}p{3in}>{\raggedright}p{1in}>{\raggedright}p{0.75in}}
\hline 
Flag  & Description  & Default value  & Required\tabularnewline
\hline 
\code{-{}-parm7}  & Path to Amber topology file  &  & required \tabularnewline
\code{-{}-rst7}  & Path to Amber coordinate file  &  & required \tabularnewline
\code{-{}-final-pdb}  & Path to final PDB file  &  & required \tabularnewline
\code{-{}-closure-list}  & List of closures  & \code{pse1 pse2}  & optional
\tabularnewline \code{-{}-tol-list}  & Tolerance list  & \code{1e-2 1e-5}  &
optional \tabularnewline \code{-{}-nproc}  & Number of processors to be used  &
\code{1}  & optional \tabularnewline \code{-{}-mpiexec}  & Parallel launcher:
e.g. \code{srun}, \code{mpirun}, \code{mpiexec}  & \code{mpiexec}  & optional
\tabularnewline \code{-{}-lm-threshold}  & LM threshold value for
placing water oxygens  & \code{0.2}  & optional \tabularnewline
\hline 
\end{tabular}
\end{table*}

Table \ref{tab:List-of-options} summarizes the required and some frequently used
parameters for \code{solv-eze}. The required flags are \code{-{}-parm7} and
\code{-{}-rst7} for the input parameter and coordinate files, as well as
\code{-{}-final-pdb}, which defines the path where the output PDB file
containing the placed water molecules will be saved. 

Closure approximations  have a significant impact on the features of the 3D
number density distribution. RISM calculations in AmberTools utilize sequential
solutions of the partial series expansion (PSE) closures. The second order PSE
closure (PSE-2) provides a balance of high-quality water placements with
reliable convergence. However, users can configure \code{solv-eze} to use other
closures with the \code{-{}-closure-list} option. Both 1D- and 3D-RISM
sequentially apply each closure, which aids convergence when higher order PSE
closures are used. For example, the sequence \code{"pse1 pse2 pse3"} can be used
to obtain solutions with the PSE-3 closure. The numerical tolerance of the
solutions is specified with \code{-{}-tol-list}, for which the second value
corresponds to the closure used for water placement, while the first value
(typically larger) applies to all preceding closures used to enhance
convergence. 

By default, waters are placed using \code{metatwist}, and the number of waters
placed depends on the \code{-{}-lm-threshold} option. If the user
wishes to try different thresholds, they can avoid repeating the 3D-RISM
calculation by rerunning \code{solv-eze} with the \code{-{}-skip-rism} flag.
The user may also select
\code{placevent}\cite{sindhikara_placevent:_2012,sindhikaraDansindPlacevent2026,sindhikaraDansindGrid2021}
or \code{gasol}\cite{fusaniOptimalWaterNetworks2018,accscAccscGAsol2026} for
water placement using the \code{-{}-backend} flag, each of which has its own optional parameters.

Finally, oxygen atoms are protonated with \code{gwh}, which attempts to optimize
the hydrogen bonding network. If only oxygen positions are desired, this step
can be omitted with \code{-{}-skip-gwh}.

By default, 3D-RISM calculations run on a single CPU core, but the user may
request parallel calculations with the \code{-{}-nproc} and \code{-{}-mpiexec}
options. 

Other options allow the user to adjust parameters of the 1D- and 3D-RISM steps
or replace them completely with precalculated files. A complete list of optional
parameters and details on advanced usage will be added to the next release of
the AmberTools manual.

\section{Validation method}

The method was evaluated by running calculations on a set of 93 protein-ligand
systems from a dataset of complexes containing bridging
waters.\cite{lu2007analysis} Protein-ligand systems with covalent ligands or
missing atoms in the binding site were omitted. Crystallographic water molecules
were considered bridging waters if they were located between $2.2$ and
$\SI{3.6}{\angstrom}$ from polar atoms (S, N or O) in both the protein and the
ligand. To identify and characterize buried bridging waters, we used the
\verb|shrake_rupley| function implemented in the \verb|MDTraj|
package\cite{McGibbon2015MDTraj}. A probe radius of \SI{1.4}{\angstrom} was used,
and a water was considered buried if its solvent-accessible surface area was
\SI{0}{\angstrom^2}.

\subsection{System Preparation and
Parameterization\label{subsec:System-Preparation-and}}

Each system (Table S1) was prepared with crystallographic waters present. First,
proteins were preprocessed with the PDBFixer application in OpenMM\cite{eastman_openmm_2017} to replace
nonstandard amino acids, add missing atoms, and protonate the system at a pH of
7. Missing residues were not modeled; termini and residues adjacent to gaps were treated in
their charged form. Histidine residues were renamed to reflect their protonation
states (HIE, HID, and HIP) using an in-house
PyMOL\cite{schrodinger_llc_pymol_2015} script.  Two copies of each system were
modeled: one to evaluate Solv-eze water placement by removing all
crystallographic waters, and the other retained crystallographic waters as a
control case to compare Solv-eze against. 

Ligand molecules were prepared using RDKit\cite{landrum_rdkit_nodate} to
protonate and correct the bond orders of organic compounds. Each small molecule
was scrutinized in its binding site environment to identify the appropriate
protonation state to be parameterized; ID and net charge for each ligand
can be found in Table S2. The small molecule ligands were then
parameterized using Antechamber 22. Atomic partial charges were derived from the
AM1-BCC\cite{jakalian_fast_2002} charge model, and the Generalized Amber Force
Field 2 (GAFF 2) \cite{he_fast_2020, vassetti_assessment_2019} was used to
describe the bonds, angles, and dihedral parameters for each compound.
Protein-ligand systems were constructed with
\code{tleap}\cite{case2023ambertools} using Amber ff19SB protein
parameters\cite{tian_ff19sb_2020}, GAFF2 ligand parameters, and standard
monatomic ion parameters appropriate for use with the SPC/E water
model\cite{berendsen_missing_1987}. After force field parameters were applied, a
system box was defined with a \SI{10}{\angstrom} minimum distance between solute
atoms and the unit cell edges, suitable for 3D-RISM calculations. 

\subsection{Water placement\label{subsec:Water-placement}}

As described in Section \ref{subsec:packaga-Water-placement}, \code{solv-eze}
sequentially runs \code{rism1d} (if necessary),
\code{rism3d.snglpnt}\cite{luchkoThreeDimensionalMolecularTheory2010,gray2022integral},
\code{metatwist}\cite{giambasu2019predicting}, and \code{gwh} from the
AmberTools software suite\cite{case2023ambertools}.

In all cases, the default \code{rism1d} result was used. For this, \code{rism1d}
solved the dielectrically consistent 1D-RISM
equations\cite{perkynsDielectricallyConsistentInteraction1992,perkynsSiteSiteTheory1992}
for pure water using the SPC/E model at a concentration of $\SI{55.34}{M}$,
dielectric constant of 78.4 and temperature of $\SI{298.15}{K}$, with a residual
tolerance of $10^{-10}$ on a grid containing 16384 points and a spacing of
$\SI{0.025}{\angstrom}$. The 1D-RISM equations were first solved with the PSE-1
closure, and the results served as an initial guess for subsequent calculations
with the PSE-2 closure. The 1D-RISM equations were solved using the modified
method of direct inversion in iterative subspace
(MDIIS)\cite{kovalenko1999solution}, utilizing 20 previous vector solutions and
an MDIIS step size of 0.3.

3D-RISM calculations with periodic boundary conditions were carried out using a
grid spacing of $\SI{0.5}{\angstrom}$. The closure chain feature was used to
obtain solutions with the PSE-1 closure, at a tolerance of $10^{-2}$, which was
used as initial guesses for the PSE-2 closure, which was solved with a tolerance
of $10^{-5}$. The MDIIS method was employed to iteratively refine the solutions,
using five previous solutions and a MDIIS step size of 0.7.

\code{metatwist} was applied in two steps to determine oxygen atom positions.
First, the Laplacian of the oxygen number density distribution from 3D-RISM was
computed with a convolution kernel of \SI{1}{\angstrom}. Then, a Laplacian blob
analysis was carried out using the same bulk concentration of water as used for
the 1D-RISM calculations, and threshold values of 0.2, 0.3, 0.5, and 0.7 were
applied. 

\code{placevent} and \code{gasol} were applied separately to place the water
oxygens using the same 3D-RISM density profile. The default options were used
for both executables (Section S1), except that spatial filtering was used for
\code{gasol}. Without spatial filtering,
\verb|gasol| places  waters throughout the simulation box, which can take more
than a day on multiple CPU cores. Thus, spatial filters were applied by
providing PDB files for the ligands and a spatial filter radius of
\SI{13.0}{\angstrom}.  In seven systems, the defined radius was reduced to avoid
overlap between the spatial filters or increased to fully encompass the ligand
while maintaining a buffer of at least \SI{3.5}{\angstrom} for water placement.
In another five cases, the spatial filters of two ligands overlapped and it was
not possible to reduce the radius while maintaining the buffer for placing the
waters. In
these cases, the ligands were merged into a single PDB file, and the radius was
increased accordingly. The details of which systems were merged and the custom
radii values are presented in Table \ref{tab:gasol-custom-systems}.

\begin{table}
\cprotect\caption{Systems that needed custom treatment with GAsol.%
\label{tab:gasol-custom-systems}}
\centering
\begin{tabular}{ccc}
\toprule 
System & Merged ligands & Radius ($\text{\AA{}}$) \tabularnewline
\midrule
 1DCP & No & 12.0 \tabularnewline 1OYF & No & 10.5 \tabularnewline 3D14 & No &
 15.0 \tabularnewline 3GCQ & Yes & 25.0 \tabularnewline 3KBZ & No & 9.5
 \tabularnewline 3T78 & Yes & 15.0 \tabularnewline 4CSV & No & 18.0
 \tabularnewline 4FU8 & Yes & 18.0 \tabularnewline 4FUJ & Yes & 15.0
 \tabularnewline 4WKC & Yes & 15.0 \tabularnewline 5IUI & No & 15.0
 \tabularnewline 6T1M & No & 15.0 \tabularnewline
\bottomrule
\end{tabular}
\end{table}

Following placement of water oxygen atoms, \code{gwh} added hydrogen atoms.
Default parameters were used.

\subsection{Minimization step\label{subsec:Minimization-step}}

To assess the stability of both 3D-RISM predicted and crystallographic water
molecules, an implicit solvent minimization step was carried out on both sets of
water molecules, allowing the water molecules to relax in their respective local
environments.  As mentioned in section \ref{subsec:System-Preparation-and},
these systems were prepared identically, except that the hydrogen positions for
the crystallographic waters were assigned using \code{tleap}, which gives the
same orientation to all water molecules. A steepest descent energy minimization
was carried out for 10,000 steps using \code{sander} with  the
Onufriev-Bashford-Case (OBC) version of the generalized Born  implicit solvent
\cite{onufriev_exploring_2004}, a surface tension of
\SI{0.007}{\kcal\per\mol\per\angstrom^2}, a salt concentration of 0, a
non-bonded distance cutoff of \SI{100}{\angstrom}, and harmonic restraints on
all non-water atoms with a \SI{100}{\kcal\per\mol\per\angstrom^2} force constant.

\subsection{Comparison with crystallographic
waters\label{subsec:comparison_w_xtal_water}}

The oxygen atoms placed with Solv-eze were compared with those identified as
bridging and buried water molecules in the crystallographic structure. The
distances between all crystallographic and predicted waters were computed and
each crystal water was uniquely paired with a predicted water such that the sum
of the separation distances was minimized using a modified Jonker-Volgenant
algorithm\cite{crouseImplementing2DRectangular2016,virtanen2020scipy}.  This
metric was defined to take into account only crystallographic water molecules,
since Solv-eze predicts additional bridging waters that are not the focus of
the present study, {\it e.g.} protein-protein bridging waters. This process was 
repeated after energy minimization.

\section{Results and Discussion}

\subsection{Effect of LM threshold on bridging water molecule
identification}
\begin{figure}
\includegraphics{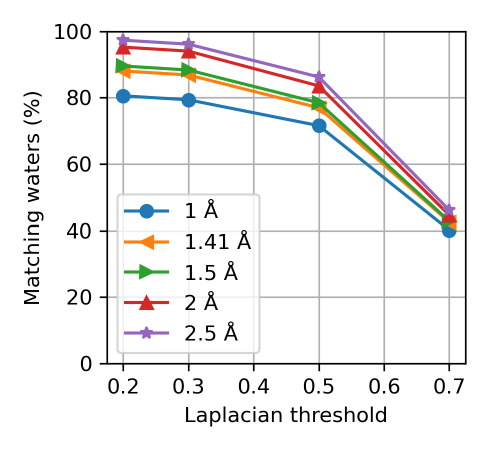}
\caption{Percentage of crystallographic water molecules matched by predicted
water molecules at each distance threshold as a function of the LM
threshold.\label{fig:metaresults}}
\end{figure}

Unlike Placevent\cite{sindhikara_placevent:_2012} and
GAsol\cite{fusaniOptimalWaterNetworks2018}, the LM water-placement method has
not been previously characterized. To assess the quality of its
predictions, we matched predicted and crystallographic water oxygens, as
described in section \ref{subsec:comparison_w_xtal_water}. In Figure
\ref{fig:metaresults}, we show the percentage of crystallographic water oxygens
matched with a predicted water oxygen within various cutoff distances, including
the van der Waals radius ($\SI{1.41}{\angstrom}$)\cite{franks2000water}. 

The van der Waals radius provides a metric based on the effective size of a
water molecule, and for LM thresholds of 0.2 and 0.3, more than 80\% of
crystal waters were matched within $\SI{1.41}{\angstrom}$ of a predicted water
and more than 90\% were within $\SI{2.00}{\angstrom}$. However, threshold values
larger than 0.3 provide poor results. 

The method plateaus as the LM threshold approaches 0.2. In our
calculations, lower threshold values were associated with increased execution
times, as expected, since the range of Laplacian values considered expands as
the threshold approaches zero. The observed trend in agreement indicates that
thresholds below 0.2 would provide only marginal gains in performance while
incurring a substantially higher computational cost. We therefore adopt a
threshold of 0.2, which offers a favorable balance between accuracy and
efficiency, and use this value throughout the remainder of the analysis.

\subsubsection{Comparison with Placevent and GAsol}

\begin{table}
\cprotect\caption{Comparison of the percentages of matching bridging waters for
different distance cutoffs using LM, Placevent and GAsol.%
\label{tab:percentages-placevent-gasol}}
\centering
\begin{tabular}{cccc}
\toprule 
Distance ($\text{\AA{}}$) & LM & Placevent & GAsol \tabularnewline
\midrule
1.00 & 80.24 & 71.86 & 47.90 \tabularnewline 1.41 & 87.72 & 79.94 & 72.75
\tabularnewline 1.50 & 89.22 & 80.84 & 74.55 \tabularnewline 2.00 & 93.41 &
88.92 & 85.93 \tabularnewline 2.50 & 95.81 & 93.71 & 91.62 \tabularnewline
\bottomrule
\end{tabular}
\end{table}

\begin{figure*}
\includegraphics{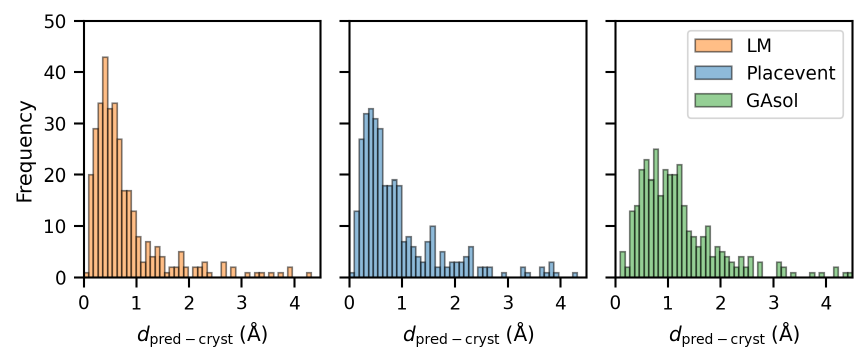}
\caption{Distributions of Solv-eze bridging waters placed at given distances
from crystallographic waters using LM, Placevent and
GAsol.\label{fig:hist-compare-methods}}
\end{figure*}

To assess the performance of LM compared to other 3D-RISM based methods, we
repeated the water placement calculations for all 93 systems using Placevent and
GAsol (Table \ref{tab:percentages-placevent-gasol}), as described in section
\ref{subsec:Water-placement}. We observe that LM achieves the best performance
for all distance thresholds, followed by Placevent, 
for which the \SI{1}{\angstrom} threshold is approximately 8
percentage points lower. GAsol performs substantially worse at the
\SI{1}{\angstrom} threshold, but improves for larger thresholds.

Placevent and LM are conceptually similar in that they both deterministically
identify high water density regions as solvation sites. Placevent places water
molecules according to density maxima, whereas LM prioritizes regions with the
most localized density. Differences arise when a smaller density peak is more
localized than a larger one: in such cases, LM places a water molecule at the
smaller peak first, whereas Placevent prioritizes the higher peak. In contrast,
GAsol uses a stochastically driven genetic algorithm to optimize the water
network.  The method uses the water density to identify possible hydration sites
and the scoring function does consider density localization, but the final water
placement depends strongly on the optimization parameters.

All three methods employ mechanisms to prevent oversolvation. LM uses Gaussian
smoothing to attenuate features smaller than $\sigma$, which may merge nearby
peaks depending on their separation and intensity. Placevent directly links the
total and local amount of solvent predicted by 3D-RISM. GAsol imposes a penalty
on water molecules placed within \SI{3}{\angstrom} from each other.

\subsubsection{Comparison with machine learning methods}

Recently, several machine learning approaches to water placement have been
developed, including HydraProt\cite{zamanosHydraProtNewDeep2024},
SuperWater\cite{kuangSuperwaterGenerativeAI2025},
GalaxyWater-CNN\cite{parkGalaxyWaterCNNPredictionWater2022}, HydraMap
v.2\cite{liHydraMapV2Prediction2023}, and
Accutar\cite{huangAccuratePredictionHydration2021}. A commonly reported metric
is the recall or coverage at a threshold of \SI{1}{\angstrom}, which is
equivalent to the fraction of water molecules matched at the same separation
used here. The results of these models vary, depending on the dataset and
runtime parameters. For example, Accutar reported \SI{1}{\angstrom} binding-site
recalls of 84.7\% on the 14-structure OppA benchmark, where waters within
\SI{4}{\angstrom} of both the protein and ligand were considered, and 62.1\% on
a 100-structure binding-site
benchmark.\cite{huangAccuratePredictionHydration2021}  Similarly, the PDBbind
core set (v2016)\cite{suComparativeAssessmentScoring2019}, which consists of
waters within \SI{5}{\angstrom} of the ligand for 285 protein-ligand complex
structures, was used to test HydraMap v.2.\cite{liHydraMapV2Prediction2023} On
this dataset, HydraMap v.1\cite{liPredictionFavorableHydration2020a} predicted
33.1\% of waters within \SI{1}{\angstrom}, versus 30.4\% for HydraMap v.2, and
42.0\% for 3D-RISM with Placevent.

On the SuperWater protein-ligand interface benchmark, where waters were defined
as being within \SI{5}{\angstrom} of both the protein surface and the bound
ligand, the authors reported precision-coverage curves rather than single
default operating points.\cite{kuangSuperwaterGenerativeAI2025} Visually
estimating the most permissive, lowest-precision points shown in the
\SI{1}{\angstrom} protein-ligand curve gives maximum coverages of approximately
66\% for SuperWater, 69\% for HydraProt, and 76\% for
GalaxyWater-CNN.\cite{kuangSuperwaterGenerativeAI2025} These  values should be
interpreted as approximate upper-end recalls on this benchmark and are not
indicative of typical performance. The trade-off for these higher recalls is
that more low-confidence waters are predicted, leading to false positives and
potentially unphysical water placement.

In several recent studies, 3D-RISM was considered as an alternative solvation
method for
comparison\cite{liHydraMapV2Prediction2023,zamanosHydraProtNewDeep2024,parkGalaxyWaterCNNPredictionWater2022}
and showed significantly lower recall than observed here. For studies that
provide complete
methods\cite{liHydraMapV2Prediction2023,parkGalaxyWaterCNNPredictionWater2022},
the major difference was the use of the Kovalenko-Hirata (KH)
closure\cite{kovalenkoPotentialMeanForce1999} instead of the PSE-2 closure used
here. The KH closure is known for its robust convergence properties, but density
peaks are damped and the excess number of solvent particles is reduced compared
to
PSE-2.\cite{joungSimpleElectrolyteSolutions2013,giambasuIonCountingExplicitSolvent2014,kast2008closed-form,caoIonDipoleCorrection3DRISM2022}
This difference likely accounts for the improved performance of Placevent
compared to past studies.

\subsection{Minimization step}

\begin{table}
\cprotect\caption{Percentage of matching bridging waters for different distance
cutoffs using a LM threshold of 0.2 from the original placement, after
energy minimization of just the predicted water molecule positions (LM), and
after energy minimization of both predicted and crystallographic water molecules
(LM \& xtal).%
\label{tab:percentages-2}}

\centering

\begin{tabular}{cccc}
\toprule 
Distance ($\text{\AA{}}$) & Original & LM & LM \& xtal\tabularnewline
\midrule
1.00 & 80.24 & 78.14 & 80.95 \tabularnewline 1.41 & 87.72 & 88.02 & 89.12
\tabularnewline 1.50 & 89.22 & 89.52 & 89.46 \tabularnewline 2.00 & 93.41 &
92.81 & 93.88 \tabularnewline 2.50 & 95.81 & 95.51 & 96.94 \tabularnewline
\bottomrule
\end{tabular}
\end{table}

\begin{figure*}
\includegraphics{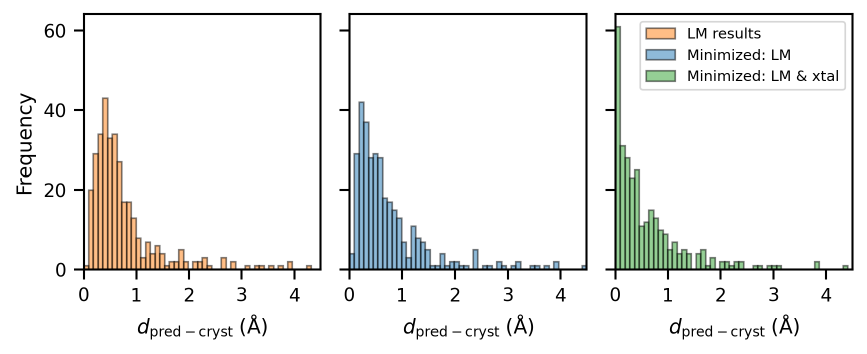}

\caption{Distributions of Solv-eze bridging waters placed at given distances
from crystallographic waters (left) before energy minimization, (center) after
energy minimizing only Solv-eze-placed waters, and (right) after energy
minimizing both Solv-eze and crystallographic waters.\label{fig:hist}}
\end{figure*}

Energy minimization is an essential step in any system preparation protocol. To
assess the effects on the placement of water molecules, we carried out energy
minimization on both crystallographic water molecules and those placed with a
LM threshold of 0.2. 

First, we minimized the energy of only the predicted waters and compared them to
the initial crystallographic structure. This led to a slight decrease in the
agreement between the predicted and crystallographic water molecules within
\SI{1}{\angstrom} (Table \ref{tab:percentages-2}). This suggests that 3D-RISM
exhibits good accuracy when compared to crystallographic data and that it also
consistently places bridging waters in accordance with the force field used.

Then, we minimized the energy of the crystallographic water molecules and
compared their positions to the energy minimized predicted waters. 
Compared to the original Solv-eze prediction and its minimized structure, we
observe a significant increase in the number of waters at
$d_{\text{pred} - \text{cryst}} \approx 0$ (last panel of Figure
\ref{fig:hist}).

The total number of bridging waters identified in the initial crystallographic
structures was 334 before minimization and 294 after. This indicates that some
crystallographic bridging waters moved away from their binding sites during
minimization. Figures S1 and S2 in the Supplementary Information show the
average distance between predicted and crystallographic waters for each system
before and after minimization, along with the number of identified bridging
waters.

Figure \ref{fig:hist} shows little change in agreement when only Solv-eze water
molecules are optimized, but we see an improvement in agreement
between the crystallographic and predicted waters after minimization. This is
primarily due to waters that already have sub-\AA{}ngstrom separations coming closer
together, indicating that the Solv-eze predictions were already close to a local
minimum for the force field.  Thus, minimization had a more pronounced effect on
the crystallographic water molecules, bringing many of these molecules closer to
the positions predicted by Solv-eze. This indicates that Solv-eze places
bridging waters in locations that are consistent with the force field and that
small deviations are likely unimportant.

When considering only buried waters, we observe better agreement between
predicted and crystallographic waters than for bridging waters in general (Table
\ref{tab:percentages-3}). After minimization, only 12 of the 159 crystallographic waters were not accounted
for within \SI{1.5}{\angstrom}. Of these, we see that 3D-RISM predicts water
density peaks at the sites of all but one water (1THN, resid 553). In another
case (3E8S, resid 418) the density peak included the crystallographic water but
the local maximum was more than \SI{1.5}{\angstrom} away. Likely reasons the
waters with nearby peaks were missed are that either the Laplacian threshold was
not met or Gaussian smoothing may have merged nearby maxima. Of the 12 waters
missed by LM, we found that Placevent without energy minimization missed only
4 (1JWB, resid 436; 1THN, resid 553; 3E8S, resid 418; 2ZVC, resid 413). This
suggests that further improvements in predicting water sites could be achieved
by adjusting the LM threshold and smoothing or combining LM and Placevent
results.

\begin{figure*}
\includegraphics{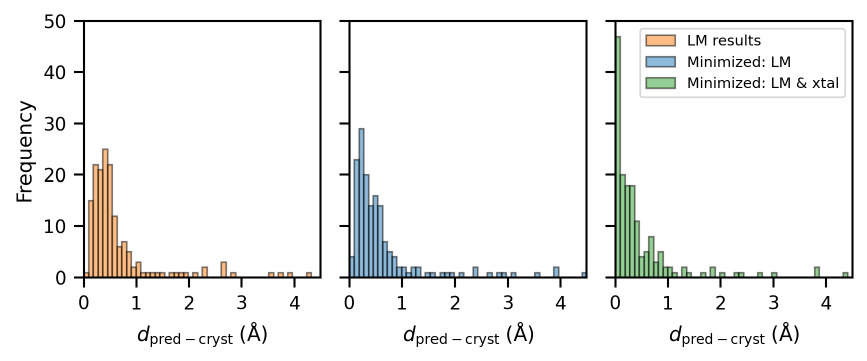}
\caption{Same as in Figure \ref{fig:hist}, but for buried
waters.\label{fig:hist2}}
\end{figure*}

We also observe that the improvement in agreement between crystallographic and
predicted waters after energy minimization is mainly for waters with
sub-\AA{}ngstrom separations (Figure \ref{fig:hist2}). This is not surprising,
as roughly half of the identified bridging waters are also buried and these
waters will be sterically confined, limiting their movement. Table
\ref{tab:percentages-3} presents the percentage of buried waters predicted
within the same thresholds as presented in Table \ref{tab:percentages-2}.

\begin{table}
\cprotect\caption{Percentage of matching buried waters for different distance
cutoffs using a LM threshold of 0.2 from the original placement, after
energy minimization of just the predicted water molecule positions (LM), and
after energy minimization of both predicted and crystallographic water molecules
(LM \& xtal).%
\label{tab:percentages-3}}
\centering
\begin{tabular}{cccc}
\toprule 
Distance ($\text{\AA{}}$) & Original & LM & LM \& xtal\tabularnewline
\midrule
1.00 & 85.19 & 85.80 & 88.68 \tabularnewline 1.41 & 89.51 & 89.51 & 91.82
\tabularnewline 1.50 & 90.12 & 89.51 & 92.45 \tabularnewline 2.00 & 92.59 &
92.59 & 94.34 \tabularnewline 2.50 & 94.44 & 94.44 & 96.23 \tabularnewline
\bottomrule
\end{tabular}
\end{table}

The total number of buried waters identified before minimization was 162 and
after minimization was 159. The difference observed in the number of buried
waters before and after minimizing the crystallographic structure is due to
small shifts in the positions of water molecules that cause their solvent-
accessible surface area to change.

\subsubsection{Water orientation}

\begin{figure}
\includegraphics{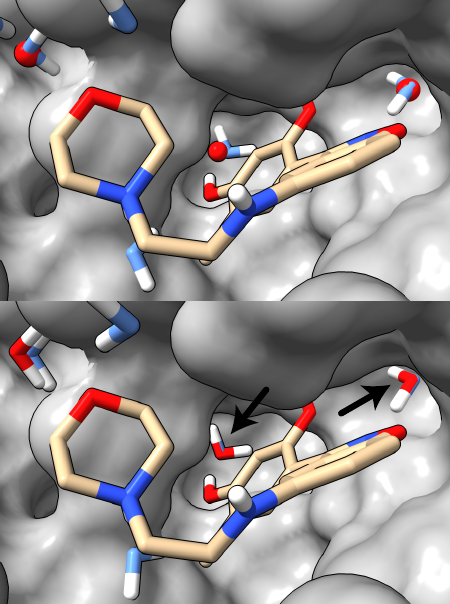}

\caption{Water molecules in the heat shock protein 90 (Hsp90)
(PDB ID: 3BMY)\cite{gopalsamyDiscoveryBenzisoxazolesPotent2008a} binding site
before (top) and after (bottom) minimization. All crystallographic waters (red)
were protonated and had the same orientation before minimization. Predicted
waters are light blue. Arrows indicate the locations of predicted waters
obscured by crystallographic waters. Other coloring as in Figure
\ref{fig:Package-workflow.}\label{fig:Water-molecules-in-binding-site}.}

\end{figure}

The orientation of a water molecule is critical for it to bridge a protein and a
ligand, as it must form hydrogen bonds with both molecules. In general, crystal
structures do not have sufficient resolution to determine the orientation of
water molecules, and we cannot directly test our predictions. Instead, we can
compare the orientations of the waters before and after minimization.

As a representative example, the heat shock protein 90 (Hsp90)
(PDB ID: 3BMY)\cite{gopalsamyDiscoveryBenzisoxazolesPotent2008a} is shown in
Figure \ref{fig:Water-molecules-in-binding-site}. Before energy minimization
(top), all three crystallographic water oxygen atoms have a predicted water
molecule within \SI{1}{\angstrom}, and the hydrogen atoms for the predicted
water molecules are determined with \code{gwh}. After minimizing the energy of
both crystallographic and predicted water molecules (bottom), we see that the
agreement between the locations of the water oxygen atoms has improved. In fact,
it is difficult to see two of the predicted waters because they overlap with the
crystallographic waters (arrows).

We also observe close agreement between the orientations predicted by GWH and
those of crystallographic waters after energy minimization. All predicted and
crystallographic water pairs in Figure \ref{fig:Water-molecules-in-binding-site}
have nearly identical orientations (arrows). As described in Section
\ref{subsec:Minimization-step}, hydrogen atoms for the crystallographic
structures were added by \code{tleap}, resulting in all water molecules having
the same orientation, regardless of their local environment. The fact that only
small changes are observed in the orientation of the predicted water molecules
and that there is agreement with the orientations of the energy minimized
crystallographic water molecules indicates that \code{gwh} provides excellent
initial guesses. Conversely, any problems with predicted water orientations are
naturally resolved through energy minimization as a standard part of system
relaxation.

\subsection{Computational requirements}

\begin{figure}
\includegraphics{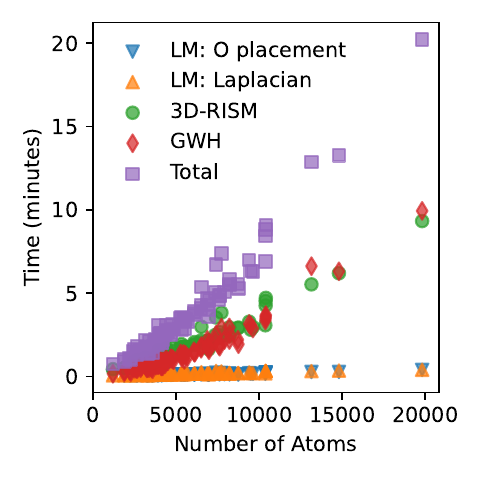}
\caption{Execution time as a function of system size. The total execution time
and the runtimes of the individual workflow components (3D-RISM, LM Laplacian
computation, LM oxygen placement, and GWH hydrogen placement) are shown for
each protein-ligand complex. \label{fig:times}}
\end{figure}

The total execution time and runtimes of the individual steps for each protein-
ligand complex are presented in Figure \ref{fig:times} as a function of the total
number of solute atoms. The only system that exceeded 15 minutes of total
execution on 12 processing cores (AMD EPYC 9654 96-Core Processor) was 3KBZ,
which has 19812 atoms and required a total of 4.67 GB of RAM. This is a modest
amount of additional time for system setup and is far less than the computation
time that would be required for exchange during molecular dynamics simulations.

Among the three water placement methods analyzed in this work, LM achieves the
best overall computational performance. For systems containing fewer than 1,500
atoms, Placevent is up to 1.5 times slower than LM. However, as the system
size increases, its runtime increases relative to LM, becoming up to four
times slower for the largest systems analyzed. Still, both methods account for
only a small fraction of the total runtime.

GAsol is the slowest of the three methods. When applied to the full simulation
box, its runtime exceeds 24 hours. Applying the spatial filter drastically
reduces the computational cost. However, with the settings used in this work,
GAsol remains substantially slower than LM. Even for the fastest system,
GAsol is 11 times slower than LM, while for the slowest system it is over 1000
times slower.

3D-RISM and \code{gwh} account for essentially all of the total execution time,
while both LM steps account for only a small portion.  While 3D-RISM scales as
$\mathcal{O}(N_\text{grid}\log N_\text{grid})$, where $N_\text{grid}$ is the
number of grid points and is proportional to the number of solute atoms in
practice, \code{gwh} scales linearly with the number of placed water atoms.
Therefore, \code{gwh} requires more time for methods that place more waters.
This is relevant for Placevent, which places 2 to 3 times as many waters as LM
in our tests. Almost all of these additional waters are in the bulk region. The
\code{gwh} step can be skipped by the user if only water oxygens are needed.

\section{Conclusions}
Modeling structures whose stability or interactions are mediated by bridging
waters requires an accurate initial placement of these molecules before
performing molecular dynamics simulations. However, experimental structural data
may be unavailable, for example, in the case of new drug candidates. In this
work, we present Solv-eze, a package that unifies three approaches for
predicting bridging waters, and propose LM as a faster method based on the
oxygen density distributions acquired from 3D-RISM calculations.

The results demonstrate that the LM method provides accurate initial water
positions when compared to crystallographic data. Using a LM threshold of
0.2, more than 80\% of the crystallographic bridging waters and
85\% of buried bridging waters were predicted within $\SI{1.00}{\angstrom}$.
After energy minimization of both the Solv-eze predictions and the experimental
positions, we observed an improvement in the overall agreement of the buried
water molecule positions, often because crystallographic water molecules
moved towards the Solv-eze predictions. Therefore, the proposed method produces
predictions consistent with the force field used and in good agreement with
crystallographic bridging water molecules, including those that may not be captured by conventional solvation with
a pre-equilibrated water box.  Although this study focuses on bridging waters at
the protein-ligand interface, the method can generate initial solvent placements
for arbitrary solutes. It should be particularly useful in cases where standard
approaches face kinetic barriers that hinder equilibration toward
the correct solvent distribution. 

\begin{acknowledgement}
This material is based upon work supported by the National Science Foundation
(NSF) under grants CHE-2102668, OAC-2320718, and OAC-2320846.  Thanks to the
American taxpayer for generously funding this research through the NIH/NIGMS
Grant R35-GM144089.  The views expressed in this paper are the responsibility of
the authors and do not necessarily reflect the official views of the NIH. We
thank David A. Case for his earlier efforts developing the water placement
protocol upon which this work was built. 
\end{acknowledgement}

\begin{suppinfo}
Figures presenting the average distance between predicted and crystallographic
waters, as well as the number of bridging waters identified for each system, are
available in the Supplementary Information. The package will be made available
free of charge as an update to AmberTools 26. Scripts and input files required
to reproduce the calculations with AmberTools 26 have been deposited in Zenodo
at DOI: 10.5281/zenodo.19684708.\cite{carvalho2026zenodo}
\end{suppinfo}

\bibliography{manuscript}

@article{McGibbon2015MDTraj,
    title = {MDTraj: A Modern Open Library for the Analysis of
    Molecular Dynamics Trajectories},
    author = {McGibbon, Robert T. and Beauchamp, Kyle A. and Harrigan,
    Matthew P. and Klein, Christoph and Swails, Jason M. and
    Hern{\'a}ndez, Carlos X.  and Schwantes, Christian R. and Wang,
    Lee-Ping and Lane, Thomas J. and Pande, Vijay S.},
    journal = {Biophysical Journal},
    volume = {109},
    number = {8},
    pages = {1528 -- 1532},
    year = {2015},
    doi = {10.1016/j.bpj.2015.08.015}
}

@Article{filipe2022molecular,
  author    = {Filipe, Hugo A. L. and Loura, Lu{\'\i}s M. S.},
  journal   = {Molecules},
  title     = {Molecular dynamics simulations: advances and applications},
  year      = {2022},
  number    = {7},
  pages     = {2105},
  volume    = {27},
  publisher = {MDPI},
  doi       = {10.3390/molecules27072105}
}

@article{bittner2024investigating,
  title={Investigating biomolecules in deep eutectic solvents with molecular dynamics simulations: current state, challenges and future perspectives},
  author={Bittner, Jan Philipp and Smirnova, Irina and Jakobtorweihen, Sven},
  journal={Molecules},
  volume={29},
  number={3},
  pages={703},
  year={2024},
  publisher={MDPI},
  doi={10.3390/molecules29030703}
}

@Article{wu2022application,
  author    = {Wu, Xiaodong and Xu, Li-Yan and Li, En-Min and Dong, Geng},
  journal   = {Chemical Biology \& Drug Design},
  title     = {Application of molecular dynamics simulation in biomedicine},
  year      = {2022},
  number    = {5},
  pages     = {789--800},
  volume    = {99},
  doi       = {10.1111/cbdd.14038},
  publisher = {Wiley Online Library},
}

@Article{wiggins1990role,
  author  = {Wiggins, Philippa M.},
  journal = {Microbiological Reviews},
  title   = {Role of water in some biological processes},
  year    = {1990},
  number  = {4},
  pages   = {432--449},
  volume  = {54},
  doi     = {10.1128/mr.54.4.432-449.1990},
}

@Article{dargaville2022water,
  author    = {Dargaville, B. L. and Hutmacher, D. W.},
  journal   = {Nature Communications},
  title     = {Water as the often neglected medium at the interface between materials and biology},
  year      = {2022},
  number    = {1},
  pages     = {4222},
  volume    = {13},
  publisher = {Nature Publishing Group UK London},
  doi       = {10.1038/s41467-022-31889-x}
}

@Article{ball2017water,
  author    = {Ball, Philip},
  journal   = {Proceedings of the National Academy of Sciences},
  title     = {Water is an active matrix of life for cell and molecular biology},
  year      = {2017},
  number    = {51},
  pages     = {13327--13335},
  volume    = {114},
  doi       = {10.1073/pnas.1703781114},
  publisher = {National Academy of Sciences},
}

@Article{bellissent2016water,
  author    = {Bellissent-Funel, Marie-Claire and Hassanali, Ali and Havenith, Martina and Henchman, Richard and Pohl, Peter and Sterpone, Fabio and Van Der Spoel, David and Xu, Yao and Garcia, Angel E.},
  journal   = {Chemical Reviews},
  title     = {Water determines the structure and dynamics of proteins},
  year      = {2016},
  number    = {13},
  pages     = {7673--7697},
  volume    = {116},
  doi       = {10.1021/acs.chemrev.5b00664},
  publisher = {ACS Publications},
}

@Article{darby2019water,
  author    = {Darby, John F. and Hopkins, Adam P. and Shimizu, Seishi and Roberts, Shirley M. and Brannigan, James A. and Turkenburg, Johan P. and Thomas, Gavin H. and Hubbard, Roderick E. and Fischer, Marcus},
  journal   = {Journal of the American Chemical Society},
  title     = {Water networks can determine the affinity of ligand binding to proteins},
  year      = {2019},
  number    = {40},
  pages     = {15818--15826},
  volume    = {141},
  doi       = {10.1021/jacs.9b06275},
  publisher = {ACS Publications},
}

@Article{zsido2025water,
  author    = {Zsid{\'o}, Bal{\'a}zs Zolt{\'a}n and Het{\'e}nyi, Csaba},
  journal   = {Expert Opinion on Drug Discovery},
  title     = {Water in drug design: pitfalls and good practices},
  year      = {2025},
  number    = {6},
  pages     = {745--764},
  volume    = {20},
  doi       = {10.1080/17460441.2025.2497912},
  publisher = {Taylor \& Francis},
}

@Article{pal2004dynamics,
  author    = {Pal, Samir Kumar and Zewail, Ahmed H.},
  journal   = {Chemical Reviews},
  title     = {Dynamics of water in biological recognition},
  year      = {2004},
  number    = {4},
  pages     = {2099--2124},
  volume    = {104},
  doi       = {10.1002/chin.200424300},
  publisher = {ACS Publications},
}

@Article{williams1994buried,
  author    = {Williams, Mark A. and Goodfellow, Julia M. and Thornton, Janet M.},
  journal   = {Protein Science},
  title     = {Buried waters and internal cavities in monomeric proteins},
  year      = {1994},
  number    = {8},
  pages     = {1224--1235},
  volume    = {3},
  doi       = {10.1002/pro.5560030808},
  publisher = {Wiley Online Library},
}

@Article{giambasu2019predicting,
  author    = {Giamba{\c{s}}u, George M. and Case, David A. and York, Darrin M.},
  journal   = {Journal of the American Chemical Society},
  title     = {Predicting site-binding modes of ions and water to nucleic acids using molecular solvation theory},
  year      = {2019},
  number    = {6},
  pages     = {2435--2445},
  volume    = {141},
  doi       = {10.1021/jacs.8b11474},
  publisher = {ACS Publications},
}

@article{caseRecentDevelopmentsAmber2025,
  title = {Recent Developments in {Amber} Biomolecular Simulations},
  author = {Case, David A. and Cerutti, David S. and Cruzeiro, Vin{\'i}cius Wilian D. and Darden, Thomas A. and Duke, Robert E. and Ghazimirsaeed, Mahdieh and Giamba{\c s}u, George M. and Giese, Timothy J. and G{\"o}tz, Andreas W. and Harris, Julie A. and Kasavajhala, Koushik and Lee, Tai-Sung and Li, Zhen and Lin, Charles and Liu, Jian and Miao, Yinglong and {Salomon-Ferrrer}, Romelia and Shen, Jana and Snyder, Ryan and Swails, Jason and Walker, Ross C. and Wang, Jinan and Wu, Xiongwu and Zeng, Jinzhe and Cheatham Iii, Thomas E. and Roe, Daniel R. and Roitberg, Adrian and Simmerling, Carlos and York, Darrin M. and Nagan, Maria C. and Merz, Kenneth M.},
  year = 2025,
  month = aug,
  journal = {Journal of Chemical Information and Modeling},
  volume = {65},
  number = {15},
  pages = {7835--7843},
  issn = {1549-9596, 1549-960X},
  doi = {10.1021/acs.jcim.5c01063},
  urldate = {2026-04-14},
  copyright = {https://creativecommons.org/licenses/by/4.0/},
  langid = {english}
}

@Article{gray2022integral,
  author    = {Gray, Jonathon G. and Giamba{\c{s}}u, George M. and Case, David A. and Luchko, Tyler},
  journal   = {The Journal of Chemical Physics},
  title     = {Integral equation models for solvent in macromolecular crystals},
  year      = {2022},
  number    = {1},
  pages     = {014801},
  volume    = {156},
  doi       = {10.1063/5.0070869},
  publisher = {AIP Publishing},
}

@book{franks2000water,
  title = {Water: A Matrix of Life},
  shorttitle = {Water},
  author = {Franks, Felix},
  year = 2000,
  month = jul,
  publisher = {The Royal Society of Chemistry},
  doi = {10.1039/9781847552341},
  urldate = {2026-04-28},
  isbn = {978-0-85404-583-9},
  langid = {english}
}

@Article{kast2008closed-form,
  author  = {Kast, Stefan M. and Kloss, Thomas},
  journal = {The Journal of Chemical Physics},
  title   = {Closed-form expressions of the chemical potential for integral equation closures with certain bridge functions},
  year    = {2008},
  month   = dec,
  number  = {23},
  pages   = {236101},
  volume  = {129},
  doi     = {10.1063/1.3041709},
  urldate = {2023-08-25},
}

@Article{roe_protocol_2020,
  author  = {Roe, Daniel R. and Brooks, Bernard R.},
  journal = {The Journal of Chemical Physics},
  title   = {A protocol for preparing explicitly solvated systems for stable molecular dynamics simulations},
  year    = {2020},
  number  = {5},
  pages   = {054123},
  volume  = {153},
  doi     = {10.1063/5.0013849},
  urldate = {2026-04-07},
}

@Article{ben-shalom_simulating_2019,
  author    = {Ben-Shalom, Ido Y. and Lin, Charles and Kurtzman, Tom and Walker, Ross C. and Gilson, Michael K.},
  journal   = {Journal of Chemical Theory and Computation},
  title     = {Simulating {Water} {Exchange} to {Buried} {Binding} {Sites}},
  year      = {2019},
  month     = apr,
  number    = {4},
  pages     = {2684--2691},
  volume    = {15},
  doi       = {10.1021/acs.jctc.8b01284},
  publisher = {American Chemical Society},
  urldate   = {2026-04-07},
}

@Article{shirtsLessonsLearnedComparing2017,
  author   = {Shirts, Michael R. and Klein, Christoph and Swails, Jason M. and Yin, Jian and Gilson, Michael K. and Mobley, David L. and Case, David A. and Zhong, Ellen D.},
  journal  = {Journal of Computer-Aided Molecular Design},
  title    = {Lessons Learned from Comparing Molecular Dynamics Engines on the {{SAMPL5}} Dataset},
  year     = {2017},
  month    = jan,
  number   = {1},
  pages    = {147--161},
  volume   = {31},
  doi      = {10.1007/s10822-016-9977-1},
  langid   = {english},
  urldate  = {2025-12-17},
}

@Article{leeCHARMMGUIInputGenerator2016,
  author    = {Lee, Jumin and Cheng, Xi and Swails, Jason M. and Yeom, Min Sun and Eastman, Peter K. and Lemkul, Justin A. and Wei, Shuai and Buckner, Joshua and Jeong, Jong Cheol and Qi, Yifei and Jo, Sunhwan and Pande, Vijay S. and Case, David A. and Brooks, Charles L. I. I. I. and MacKerell, Alexander D. Jr. and Klauda, Jeffery B. and Im, Wonpil},
  journal   = {Journal of Chemical Theory and Computation},
  title     = {{{CHARMM-GUI Input Generator}} for {{NAMD}}, {{GROMACS}}, {{AMBER}}, {{OpenMM}}, and {{CHARMM}}/{{OpenMM}} Simulations Using the {{CHARMM36}} Additive Force Field},
  year      = {2016},
  month     = jan,
  number    = {1},
  pages     = {405--413},
  volume    = {12},
  doi       = {10.1021/acs.jctc.5b00935},
  publisher = {American Chemical Society},
  urldate   = {2026-04-14},
}

@misc{ParmEdParmEd2025,
  title = {{{ParmEd}}/{{ParmEd}}},
  author = {Swails, Jason M.},
  year = 2025,
  month = dec,
  urldate = {2025-12-17},
  howpublished = {ParmEd},
  url = {https://github.com/ParmEd/ParmEd},
}

@Article{roePTRAJCPPTRAJSoftware2013,
  author     = {Roe, Daniel R. and Cheatham, Thomas E.},
  journal    = {Journal of Chemical Theory and Computation},
  title      = {{{PTRAJ}} and {{CPPTRAJ}}: Software for Processing and Analysis of Molecular Dynamics Trajectory Data},
  year       = {2013},
  month      = jul,
  number     = {7},
  pages      = {3084--3095},
  volume     = {9},
  doi        = {10.1021/ct400341p},
  shorttitle = {Ptraj and Cpptraj},
  urldate    = {2014-11-10},
}

@Article{luchkoThreeDimensionalMolecularTheory2010,
  author  = {Luchko, Tyler and Gusarov, Sergey and Roe, Daniel R. and Simmerling, Carlos and Case, David A. and Tuszynski, Jack and Kovalenko, Andriy},
  journal = {Journal of Chemical Theory and Computation},
  title   = {Three-Dimensional Molecular Theory of Solvation Coupled with Molecular Dynamics in {Amber}},
  year    = {2010},
  month   = mar,
  number  = {3},
  pages   = {607--624},
  volume  = {6},
  doi     = {10.1021/ct900460m},
  pmcid   = {PMC2861832},
  pmid    = {20440377},
  urldate = {2014-08-29},
}

@Article{perkynsDielectricallyConsistentInteraction1992,
  author  = {Perkyns, J. S. and Montgomery Pettitt, B.},
  journal = {Chemical Physics Letters},
  title   = {A Dielectrically Consistent Interaction Site Theory for Solvent---Electrolyte Mixtures},
  year    = {1992},
  month   = mar,
  number  = {6},
  pages   = {626--630},
  volume  = {190},
  doi     = {10.1016/0009-2614(92)85201-K},
  langid  = {english},
  urldate = {2020-01-13},
}

@Article{perkynsSiteSiteTheory1992,
  author  = {Perkyns, John and Pettitt, B. Montgomery},
  journal = {The Journal of Chemical Physics},
  title   = {A Site--Site Theory for Finite Concentration Saline Solutions},
  year    = {1992},
  number  = {10},
  pages   = {7656--7666},
  volume  = {97},
  doi     = {10.1063/1.463485},
  urldate = {2014-09-15},
}

@Misc{schrodinger_llc_pymol_2015,
  author = {Schrodinger, L. L. C.},
  month  = nov,
  title  = {The PyMOL Molecular Graphics System, Version 3.1},
  year   = {2024},
}

@Article{eastman_openmm_2017,
  author     = {Eastman, Peter and Swails, Jason and Chodera, John D. and McGibbon, Robert T. and Zhao, Yutong and Beauchamp, Kyle A. and Wang, Lee-Ping and Simmonett, Andrew C. and Harrigan, Matthew P. and Stern, Chaya D. and Wiewiora, Rafal P. and Brooks, Bernard R. and Pande, Vijay S.},
  journal    = {Plos Computational Biology},
  title      = {{OpenMM} 7: {Rapid} development of high performance algorithms for molecular dynamics},
  year       = {2017},
  month      = jul,
  number     = {7},
  pages      = {e1005659},
  volume     = {13},
  doi        = {10.1371/journal.pcbi.1005659},
  language   = {en},
  shorttitle = {{OpenMM} 7},
  urldate    = {2026-04-06},
}

@Misc{landrum_rdkit_nodate,
  author  = {Landrum, Greg},
  title   = {{RDKit}: {Open}-source cheminformatics},
  year    = {2025},
  doi     = {10.5281/zenodo.16996017},
  urldate = {2025-08-30},
  url     = {https://www.rdkit.org}
}

@Article{jakalian_fast_2002,
  author     = {Jakalian, Araz and Jack, David B. and Bayly, Christopher I.},
  journal    = {Journal of Computational Chemistry},
  title      = {Fast, efficient generation of high-quality atomic charges. {AM1}-{BCC} model: {II}. {Parameterization} and validation},
  year       = {2002},
  month      = dec,
  number     = {16},
  pages      = {1623--1641},
  volume     = {23},
  doi        = {10.1002/jcc.10128},
  language   = {eng},
  pmid       = {12395429},
  shorttitle = {Fast, efficient generation of high-quality atomic charges. {AM1}-{BCC} model},
}

@Article{he_fast_2020,
  author  = {He, Xibing and Man, Viet H. and Yang, Wei and Lee, Tai-Sung and Wang, Junmei},
  journal = {The Journal of Chemical Physics},
  title   = {A fast and high-quality charge model for the next generation general {AMBER} force field},
  year    = {2020},
  month   = sep,
  number  = {11},
  pages   = {114502},
  volume  = {153},
  doi     = {10.1063/5.0019056},
  pmcid   = {PMC7728379},
  pmid    = {32962378},
  urldate = {2026-04-06},
}

@Article{vassetti_assessment_2019,
  author  = {Vassetti, Dario and Pagliai, Marco and Procacci, Piero},
  journal = {Journal of Chemical Theory and Computation},
  title   = {Assessment of {GAFF2} and {OPLS}-{AA} {General} {Force} {Fields} in {Combination} with the {Water} {Models} {TIP3P}, {SPCE}, and {OPC3} for the {Solvation} {Free} {Energy} of {Druglike} {Organic} {Molecules}},
  year    = {2019},
  month   = mar,
  number  = {3},
  pages   = {1983--1995},
  volume  = {15},
  doi     = {10.1021/acs.jctc.8b01039},
  urldate = {2026-04-06},
}

@Article{tian_ff19sb_2020,
  author     = {Tian, Chuan and Kasavajhala, Koushik and Belfon, Kellon A. A. and Raguette, Lauren and Huang, He and Migues, Angela N. and Bickel, John and Wang, Yuzhang and Pincay, Jorge and Wu, Qin and Simmerling, Carlos},
  journal    = {Journal of Chemical Theory and Computation},
  title      = {{ff19SB}: {Amino}-{Acid}-{Specific} {Protein} {Backbone} {Parameters} {Trained} against {Quantum} {Mechanics} {Energy} {Surfaces} in {Solution}},
  year       = {2020},
  month      = jan,
  number     = {1},
  pages      = {528--552},
  volume     = {16},
  doi        = {10.1021/acs.jctc.9b00591},
  shorttitle = {{ff19SB}},
  urldate    = {2026-04-06},
}

@Article{berendsen_missing_1987,
  author  = {Berendsen, H. J. C. and Grigera, J. R. and Straatsma, T. P.},
  journal = {Journal of Physical Chemistry},
  title   = {The missing term in effective pair potentials},
  year    = {1987},
  month   = nov,
  number  = {24},
  pages   = {6269--6271},
  volume  = {91},
  doi     = {10.1021/j100308a038},
  urldate = {2026-04-06},
}

@Article{chandler_optimized_1972,
  author       = {Chandler, David and Andersen, Hans C.},
  journal      = {The Journal of Chemical Physics},
  title        = {Optimized Cluster Expansions for Classical Fluids. {II}. Theory of Molecular Liquids},
  year         = {1972},
  month        = sep,
  number       = {5},
  pages        = {1930--1937},
  volume       = {57},
  doi          = {10.1063/1.1678513},
  shortjournal = {J. Chem. Phys.},
  urldate      = {2026-04-07},
}

@Article{hirata_extended_1981,
  author       = {Hirata, Fumio and Rossky, Peter J.},
  journal      = {Chemical Physics Letters},
  title        = {An extended rism equation for molecular polar fluids},
  year         = {1981},
  month        = oct,
  number       = {2},
  pages        = {329--334},
  volume       = {83},
  doi          = {10.1016/0009-2614(81)85474-7},
  shortjournal = {Chemical Physics Letters},
  urldate      = {2026-04-07},
}

@InCollection{hirata_three-dimensional_2004,
  author    = {Kovalenko, Andriy},
  booktitle = {Molecular Theory of Solvation},
  publisher = {Kluwer Academic Publishers},
  title     = {Three-dimensional Rism Theory for Molecular Liquids and Solid-Liquid Interfaces},
  year      = {2004},
  address   = {Dordrecht},
  editor    = {Hirata, Fumio},
  isbn      = {978-1-4020-1562-5},
  pages     = {169--275},
  volume    = {24},
  doi       = {10.1007/1-4020-2590-4_4},
  langid    = {english},
  urldate   = {2026-04-07},
}

@Article{kovalenko_three-dimensional_1998,
  author       = {Kovalenko, Andriy and Hirata, Fumio},
  journal      = {Chemical Physics Letters},
  title        = {Three-dimensional density profiles of water in contact with a solute of arbitrary shape: a {RISM} approach},
  year         = {1998},
  month        = jun,
  number       = {1},
  pages        = {237--244},
  volume       = {290},
  doi          = {10.1016/S0009-2614(98)00471-0},
  shortjournal = {Chemical Physics Letters},
  shorttitle   = {Three-dimensional density profiles of water in contact with a solute of arbitrary shape},
  urldate      = {2026-04-07},
}

@Article{widom_topics_1963,
  author       = {Widom, B.},
  journal      = {The Journal of Chemical Physics},
  title        = {Some Topics in the Theory of Fluids},
  year         = {1963},
  month        = dec,
  number       = {11},
  pages        = {2808--2812},
  volume       = {39},
  doi          = {10.1063/1.1734110},
  langid       = {english},
  publisher    = {{AIP} Publishing},
  shortjournal = {J. Chem. Phys.},
  urldate      = {2026-04-07},
}

@Article{perego_molecular_2015,
  author       = {Perego, C. and Salvalaglio, M. and Parrinello, M.},
  journal      = {The Journal of Chemical Physics},
  title        = {Molecular dynamics simulations of solutions at constant chemical potential},
  year         = {2015},
  month        = apr,
  number       = {14},
  pages        = {144113},
  volume       = {142},
  doi          = {10.1063/1.4917200},
  shortjournal = {J. Chem. Phys.},
  urldate      = {2026-04-07},
}

@Article{ge_enhancing_2022,
  author       = {Ge, Yunhui and Melling, Oliver J. and Dong, Weiming and Essex, Jonathan W. and Mobley, David L.},
  journal      = {Journal of Computer-Aided Molecular Design},
  title        = {Enhancing sampling of water rehydration upon ligand binding using variants of grand canonical Monte Carlo},
  year         = {2022},
  month        = oct,
  number       = {10},
  pages        = {767--779},
  volume       = {36},
  doi          = {10.1007/s10822-022-00479-w},
  langid       = {english},
  shortjournal = {J Comput Aided Mol Des},
  urldate      = {2026-04-07},
}

@Article{deng_efficient_2024,
  author       = {Deng, Jiahua and Cui, Qiang},
  journal      = {Journal of Chemical Theory and Computation},
  title        = {Efficient Sampling of Cavity Hydration in Proteins with Nonequilibrium Grand Canonical Monte Carlo and Polarizable Force Fields},
  year         = {2024},
  month        = mar,
  number       = {5},
  pages        = {1897--1911},
  volume       = {20},
  doi          = {10.1021/acs.jctc.4c00013},
  publisher    = {American Chemical Society},
  shortjournal = {J. Chem. Theory Comput.},
  urldate      = {2026-04-07},
}

@Article{samways_grand_2020,
  author       = {Samways, Marley L. and Bruce Macdonald, Hannah E. and Essex, Jonathan W.},
  journal      = {Journal of Chemical Information and Modeling},
  title        = {grand: A Python Module for Grand Canonical Water Sampling in {OpenMM}},
  year         = {2020},
  month        = oct,
  number       = {10},
  pages        = {4436--4441},
  volume       = {60},
  doi          = {10.1021/acs.jcim.0c00648},
  publisher    = {American Chemical Society},
  shortjournal = {J. Chem. Inf. Model.},
  shorttitle   = {grand},
  urldate      = {2026-04-07},
}

@Article{braun_best_2019,
  author   = {Braun, Efrem and Gilmer, Justin and Mayes, Heather B. and Mobley, David L. and Monroe, Jacob I. and Prasad, Samarjeet and Zuckerman, Daniel M.},
  journal  = {Living Journal of Computational Molecular Science},
  title    = {Best Practices for Foundations in Molecular Simulations [Article v1.0]},
  year     = {2019},
  number   = {1},
  pages    = {5957--5957},
  volume   = {1},
  doi      = {10.33011/livecoms.1.1.5957},
  langid   = {english},
  rights   = {Copyright (c) 2018},
  urldate  = {2026-04-07},
}

@Article{birkhoff_proof_1931,
  author    = {Birkhoff, George D.},
  journal   = {Proceedings of the National Academy of Sciences},
  title     = {Proof of the {Ergodic} {Theorem}},
  year      = {1931},
  month     = dec,
  number    = {12},
  pages     = {656--660},
  volume    = {17},
  doi       = {10.1073/pnas.17.2.656},
  publisher = {Proceedings of the National Academy of Sciences},
  urldate   = {2026-04-08},
}

@Article{lu2007analysis,
  author    = {Lu, Yipin and Wang, Renxiao and Yang, Chao-Yie and Wang, Shaomeng},
  journal   = {Journal of Chemical Information and Modeling},
  title     = {Analysis of ligand-bound water molecules in high-resolution crystal structures of protein- ligand complexes},
  year      = {2007},
  number    = {2},
  pages     = {668--675},
  volume    = {47},
  doi       = {10.1021/ci6003527},
  publisher = {ACS Publications},
}

@Article{case2023ambertools,
  author       = {Case, David A. and Aktulga, Hasan Metin and Belfon, Kellon and Cerutti, David S. and Cisneros, G. Andr{\'e}s and Cruzeiro, Vin{\'i}cius Wilian D. and Forouzesh, Negin and Giese, Timothy J. and G{\"o}tz, Andreas W. and Gohlke, Holger and Izadi, Saeed and Kasavajhala, Koushik and Kaymak, Mehmet C. and King, Edward and Kurtzman, Tom and Lee, Tai-Sung and Li, Pengfei and Liu, Jian and Luchko, Tyler and Luo, Ray and Manathunga, Madushanka and Machado, Matias R. and Nguyen, Hai Minh and O'Hearn, Kurt A. and Onufriev, Alexey V. and Pan, Feng and Pantano, Sergio and Qi, Ruxi and Rahnamoun, Ali and Risheh, Ali and Schott-Verdugo, Stephan and Shajan, Akhil and Swails, Jason and Wang, Junmei and Wei, Haixin and Wu, Xiongwu and Wu, Yongxian and Zhang, Shi and Zhao, Shiji and Zhu, Qiang and Cheatham, Thomas E. I. I. I. and Roe, Daniel R. and Roitberg, Adrian and Simmerling, Carlos and York, Darrin M. and Nagan, Maria C. and Merz, Kenneth M. Jr.},
  journal      = {Journal of Chemical Information and Modeling},
  title        = {{AmberTools}},
  year         = {2023},
  month        = oct,
  number       = {20},
  pages        = {6183--6191},
  volume       = {63},
  date         = {2023-10-23},
  doi          = {10.1021/acs.jcim.3c01153},
  publisher    = {American Chemical Society},
  shortjournal = {J. Chem. Inf. Model.},
  urldate      = {2023-10-11},
}

@Article{Levy2006,
  author       = {Levy, Yaakov and Onuchic, Jos{\'e} N.},
  title        = {Water mediation in protein folding and molecular recognition},
  issn         = {1936-122X, 1936-1238},
  number       = {Volume 35, 2006},
  pages        = {389--415},
  volume       = {35},
  date         = {2006-06},
  year         = {2006},
  doi          = {10.1146/annurev.biophys.35.040405.102134},
  journal      = {Annual Review of Biophysics},
  langid       = {english},
  publisher    = {Annual Reviews},
  urldate      = {2026-04-22},
}

@Article{kovalenko1999solution,
  author    = {Kovalenko, Andriy and Ten-no, Seiichiro and Hirata, Fumio},
  journal   = {Journal of Computational Chemistry},
  title     = {Solution of three-dimensional reference interaction site model and hypernetted chain equations for simple point charge water by modified method of direct inversion in iterative subspace},
  year      = {1999},
  number    = {9},
  pages     = {928--936},
  volume    = {20},
  copyright = {Copyright (c) 1999 John Wiley \& Sons, Inc.},
  doi       = {10.1002/(SICI)1096-987X(19990715)20:9<928::AID-JCC4>3.0.CO;2-X},
  language  = {en},
  urldate   = {2023-08-16},
}

@Article{abraham_gromacs_2015,
  author     = {Abraham, Mark James and Murtola, Teemu and Schulz, Roland and P{\'a}ll, Szil{\'a}rd and Smith, Jeremy C. and Hess, Berk and Lindahl, Erik},
  journal    = {SoftwareX},
  title      = {{GROMACS}: {High} performance molecular simulations through multi-level parallelism from laptops to supercomputers},
  year       = {2015},
  month      = sep,
  pages      = {19--25},
  volume     = {1-2},
  doi        = {10.1016/j.softx.2015.06.001},
  shorttitle = {{GROMACS}},
  urldate    = {2026-04-07},
}

@Article{hedgesBioSimSpaceInteroperablePython2019,
  author     = {Hedges, Lester O. and Mey, Antonia S.J.S. and Laughton, Charles A. and Gervasio, Francesco L. and Mulholland, Adrian J. and Woods, Christopher J. and Michel, Julien},
  journal    = {Journal of Open Source Software},
  title      = {{{BioSimSpace}}: {{An}} Interoperable {{Python}} Framework for Biomolecular Simulation},
  year       = {2019},
  month      = nov,
  number     = {43},
  pages      = {1831},
  volume     = {4},
  doi        = {10.21105/joss.01831},
  langid     = {english},
  shorttitle = {{{BioSimSpace}}},
  urldate    = {2026-04-14},
}

@Article{onufriev_exploring_2004,
  author    = {Onufriev, Alexey and Bashford, Donald and Case, David A.},
  journal   = {Proteins: Structure, Function, and Bioinformatics},
  title     = {Exploring protein native states and large-scale conformational changes with a modified generalized born model},
  year      = {2004},
  number    = {2},
  pages     = {383--394},
  volume    = {55},
  copyright = {Copyright (c) 2004 Wiley-Liss, Inc.},
  doi       = {10.1002/prot.20033},
  language  = {en},
  urldate   = {2026-04-06},
}

@Article{beglov_integral_1997,
  author       = {Beglov, Dmitrii and Roux, Beno{\^i}t},
  journal      = {Journal of Physical Chemistry B},
  title        = {An Integral Equation To Describe the Solvation of Polar Molecules in Liquid Water},
  year         = {1997},
  month        = sep,
  number       = {39},
  pages        = {7821--7826},
  volume       = {101},
  doi          = {10.1021/jp971083h},
  publisher    = {American Chemical Society},
  shortjournal = {J. Phys. Chem. B},
  urldate      = {2026-04-07},
}

@Article{mezei_cavity-biased_1980,
  author    = {Mezei, Mihaly},
  journal   = {Molecular Physics},
  title     = {A cavity-biased (T, V, {$\mu$}) Monte Carlo method for the computer simulation of fluids},
  year      = {1980},
  month     = jul,
  number    = {4},
  pages     = {901--906},
  volume    = {40},
  doi       = {10.1080/00268978000101971},
  publisher = {Taylor \& Francis},
  urldate   = {2026-04-07},
}

@Article{wall_biomolecular_2019,
  author    = {Wall, Michael E. and Calabr{\'o}, Gaetano and Bayly, Christopher I. and Mobley, David L. and Warren, Gregory L.},
  journal   = {Journal of the American Chemical Society},
  title     = {Biomolecular {Solvation} {Structure} {Revealed} by {Molecular} {Dynamics} {Simulations}},
  year      = {2019},
  month     = mar,
  number    = {11},
  pages     = {4711--4720},
  volume    = {141},
  doi       = {10.1021/jacs.8b13613},
  publisher = {American Chemical Society},
  urldate   = {2026-04-21},
}

@Article{joCHARMMGUIWebbasedGraphical2008,
  author       = {Jo, Sunhwan and Kim, Taehoon and Iyer, Vidyashankara G. and Im, Wonpil},
  journal      = {Journal of Computational Chemistry},
  title        = {{{CHARMM-GUI}}: {{A}} Web-Based Graphical User Interface for {{CHARMM}}},
  year         = {2008},
  number       = {11},
  pages        = {1859--1865},
  volume       = {29},
  date         = {2008},
  doi          = {10.1002/jcc.20945},
  langid       = {english},
  rights       = {Copyright (c) 2008 Wiley Periodicals, Inc.},
  shorttitle   = {{{CHARMM-GUI}}},
  urldate      = {2026-04-14},
}

@Misc{carvalho2026zenodo,
  author       = {Carvalho, Felipe Silva and Ramsey, Steven and Kurtzman, Tom and Luchko, Tyler},
  title        = {Replication package for "{S}olv-eze: {A}utomated Placement of Explicit Water Molecules Using {3D-RISM}},
  year         = {2026},
  publisher    = {Zenodo},
  doi          = {10.5281/zenodo.19684708},
  url          = {https://doi.org/10.5281/zenodo.19684708}
}

@article{sindhikara_placevent:_2012,
	title = {Placevent: An algorithm for prediction of explicit solvent atom distribution-Application to {HIV}-1 protease and F-{ATP} synthase},
	volume = {33},
	issn = {1096-987X},
	url = {http://onlinelibrary.wiley.com/doi/10.1002/jcc.22984/abstract},
	doi = {10.1002/jcc.22984},
	shorttitle = {Placevent},
	pages = {1536--1543},
	number = {18},
	journal = {Journal of Computational Chemistry},
	shortjournal = {J. Comput. Chem.},
	author = {Sindhikara, Daniel J. and Yoshida, Norio and Hirata, Fumio},
	urldate = {2018-02-27},
	date = {2012-07-05},
	langid = {english},
	year = {2012}
}

@article{huangAccuratePredictionHydration2021,
  title = {Accurate {{Prediction}} of {{Hydration Sites}} of {{Proteins Using Energy Model With Atom Embedding}}},
  author = {Huang, Pin and Xing, Haoming and Zou, Xun and Han, Qi and Liu, Ke and Sun, Xiangyan and Wu, Junqiu and Fan, Jie},
  year = 2021,
  month = sep,
  journal = {Frontiers in Molecular Biosciences},
  volume = {8},
  publisher = {Frontiers},
  issn = {2296-889X},
  doi = {10.3389/fmolb.2021.756075},
  urldate = {2026-06-19},
  langid = {english}
}

@article{kuangSuperwaterGenerativeAI2025,
  title = {Superwater as a Generative {{AI}} Framework to Predict Water Molecule Positions on Protein Structures},
  author = {Kuang, Xiaohan and Liu, Yunchao Lance and Lin, Xiaobo and {Spencer-Smith}, Jesse and Derr, Tyler and Wu, Yinghao and Bitter, Hans and Hu, Yongbo and Meiler, Jens and Su, Zhaoqian},
  year = 2025,
  month = dec,
  journal = {Communications Chemistry},
  volume = {8},
  number = {1},
  pages = {397},
  publisher = {Nature Publishing Group},
  issn = {2399-3669},
  doi = {10.1038/s42004-025-01789-4},
  urldate = {2026-06-18},
  copyright = {2025 The Author(s)},
  langid = {english}
}

@article{parkGalaxyWaterCNNPredictionWater2022,
  title = {{{GalaxyWater-CNN}}: {{Prediction}} of {{Water Positions}} on the {{Protein Structure}} by a {{3D-Convolutional Neural Network}}},
  shorttitle = {{{GalaxyWater-CNN}}},
  author = {Park, Sangwoo and Seok, Chaok},
  year = 2022,
  month = jul,
  journal = {Journal of Chemical Information and Modeling},
  volume = {62},
  number = {13},
  pages = {3157--3168},
  publisher = {American Chemical Society},
  issn = {1549-9596},
  doi = {10.1021/acs.jcim.2c00306},
  urldate = {2026-06-18}
}

@article{zamanosHydraProtNewDeep2024,
  title = {{{HydraProt}}: {{A New Deep Learning Tool}} for {{Fast}} and {{Accurate Prediction}} of {{Water Molecule Positions}} for {{Protein Structures}}},
  shorttitle = {{{HydraProt}}},
  author = {Zamanos, Andreas and Ioannakis, George and Emiris, Ioannis Z.},
  year = 2024,
  month = apr,
  journal = {Journal of Chemical Information and Modeling},
  volume = {64},
  number = {7},
  pages = {2594--2611},
  publisher = {American Chemical Society},
  issn = {1549-9596},
  doi = {10.1021/acs.jcim.3c01559},
  urldate = {2025-03-28}
}

@article{fusaniOptimalWaterNetworks2018,
  title = {Optimal Water Networks in Protein Cavities with {{GAsol}} and {{3D-RISM}}},
  author = {Fusani, Lucia and Wall, Ian and Palmer, David and Cortes, Alvaro},
  year = 2018,
  month = jun,
  journal = {Bioinformatics},
  volume = {34},
  number = {11},
  pages = {1947--1948},
  issn = {1367-4803},
  doi = {10.1093/bioinformatics/bty024},
  urldate = {2022-04-12}
}

@article{baderAtomsMolecules1985,
  title = {Atoms in Molecules},
  author = {Bader, R. F. W.},
  year = 1985,
  month = jan,
  journal = {Accounts of Chemical Research},
  volume = {18},
  number = {1},
  pages = {9--15},
  issn = {0001-4842, 1520-4898},
  doi = {10.1021/ar00109a003},
  urldate = {2026-06-22},
  langid = {english}
}

@article{lindebergFeatureDetectionAutomatic1998,
  title = {Feature {{Detection}} with {{Automatic Scale Selection}}},
  author = {Lindeberg, Tony},
  year = 1998,
  month = nov,
  journal = {International Journal of Computer Vision},
  volume = {30},
  number = {2},
  pages = {79--116},
  issn = {1573-1405},
  doi = {10.1023/A:1008045108935},
  urldate = {2026-06-22},
  langid = {english}
}

@misc{accscAccscGAsol2026,
  title = {{GAsol}},
  author = {Cortes, Alvaro},
  year = 2026,
  month = feb,
  urldate = {2026-06-30},
  copyright = {BSD-3-Clause},
  url = {https://github.com/accsc/GAsol}
}

@misc{sindhikaraDansindGrid2021,
  title = {Grid},
  author = {Sindhikara, Daniel},
  year = 2021,
  month = aug,
  urldate = {2026-06-30},
  copyright = {LGPL-3.0},
  url = {https://github.com/dansind/grid}
}

@misc{sindhikaraDansindPlacevent2026,
  title = {{Placevent}},
  author = {Sindhikara, Daniel},
  year = 2026,
  month = feb,
  urldate = {2026-06-30},
  copyright = {GPL-3.0},
  url = {https://github.com/dansind/Placevent}
}

@article{liHydraMapV2Prediction2023,
  title = {{{HydraMap}} v.2: {{Prediction}} of {{Hydration Sites}} and {{Desolvation Energy}} with {{Refined Statistical Potentials}}},
  shorttitle = {{{HydraMap}} v.2},
  author = {Li, Yan and Zhang, Zhe and Wang, Renxiao},
  year = 2023,
  month = aug,
  journal = {Journal of Chemical Information and Modeling},
  volume = {63},
  number = {15},
  pages = {4749--4761},
  publisher = {American Chemical Society},
  issn = {1549-9596},
  doi = {10.1021/acs.jcim.3c00408},
  urldate = {2026-06-30}
}

@article{suComparativeAssessmentScoring2019,
  title = {Comparative {{Assessment}} of {{Scoring Functions}}: {{The CASF-2016 Update}}},
  shorttitle = {Comparative {{Assessment}} of {{Scoring Functions}}},
  author = {Su, Minyi and Yang, Qifan and Du, Yu and Feng, Guoqin and Liu, Zhihai and Li, Yan and Wang, Renxiao},
  year = 2019,
  month = feb,
  journal = {Journal of Chemical Information and Modeling},
  volume = {59},
  number = {2},
  pages = {895--913},
  publisher = {American Chemical Society},
  issn = {1549-9596},
  doi = {10.1021/acs.jcim.8b00545},
  urldate = {2026-06-30}
}

@article{liPredictionFavorableHydration2020a,
  title = {Prediction of the {{Favorable Hydration Sites}} in a {{Protein Binding Pocket}} and {{Its Application}} to {{Scoring Function Formulation}}},
  author = {Li, Yan and Gao, Yingduo and Holloway, M. Katharine and Wang, Renxiao},
  year = 2020,
  month = sep,
  journal = {Journal of Chemical Information and Modeling},
  volume = {60},
  number = {9},
  pages = {4359--4375},
  publisher = {American Chemical Society},
  issn = {1549-9596},
  doi = {10.1021/acs.jcim.9b00619},
  urldate = {2026-06-30}
}

@article{kovalenkoPotentialMeanForce1999,
  title = {Potential of {{Mean Force}} between {{Two Molecular Ions}} in a {{Polar Molecular Solvent}}: {{A Study}} by the {{Three-Dimensional Reference Interaction Site Model}}},
  shorttitle = {Potential of {{Mean Force}} between {{Two Molecular Ions}} in a {{Polar Molecular Solvent}}},
  author = {Kovalenko, Andriy and Hirata, Fumio},
  year = 1999,
  month = sep,
  journal = {The Journal of Physical Chemistry B},
  volume = {103},
  number = {37},
  pages = {7942--7957},
  publisher = {American Chemical Society},
  issn = {1520-6106},
  doi = {10.1021/jp991300+},
  urldate = {2021-03-24}
}

@article{giambasuIonCountingExplicitSolvent2014,
  title = {Ion {{Counting}} from {{Explicit-Solvent Simulations}} and {{3D-RISM}}},
  author = {Giamba{\c s}u, George M. and Luchko, Tyler and Herschlag, Daniel and York, Darrin M. and Case, David A.},
  year = 2014,
  month = feb,
  journal = {Biophysical Journal},
  volume = {106},
  number = {4},
  pages = {883--894},
  issn = {0006-3495},
  doi = {10.1016/j.bpj.2014.01.021},
  urldate = {2014-02-25},
  pmcid = {PMC3944826},
  pmid = {24559991}
}

@article{joungSimpleElectrolyteSolutions2013,
  title = {Simple Electrolyte Solutions: {{Comparison}} of {{DRISM}} and Molecular Dynamics Results for Alkali Halide Solutions},
  shorttitle = {Simple Electrolyte Solutions},
  author = {Joung, In Suk and Luchko, Tyler and Case, David A.},
  year = 2013,
  month = jan,
  journal = {The Journal of Chemical Physics},
  volume = {138},
  number = {4},
  pages = {044103},
  issn = {00219606},
  doi = {doi:10.1063/1.4775743},
  urldate = {2013-10-14}
}

@article{caoIonDipoleCorrection3DRISM2022,
  title = {The {{Ion-Dipole Correction}} of the {{3DRISM Solvation Model}} to {{Accurately Compute Water Distributions}} around {{Negatively Charged Biomolecules}}},
  author = {Cao, Siqin and Qiu, Yunrui and Unarta, Ilona C. and Goonetilleke, Eshani C. and Huang, Xuhui},
  year = 2022,
  month = nov,
  journal = {The Journal of Physical Chemistry B},
  volume = {126},
  number = {43},
  pages = {8632--8645},
  publisher = {American Chemical Society},
  issn = {1520-6106},
  doi = {10.1021/acs.jpcb.2c04431},
  urldate = {2026-01-12}
}

@article{crouseImplementing2DRectangular2016,
  title = {On Implementing {{2D}} Rectangular Assignment Algorithms},
  author = {Crouse, David F.},
  year = 2016,
  month = aug,
  journal = {IEEE Transactions on Aerospace and Electronic Systems},
  volume = {52},
  number = {4},
  pages = {1679--1696},
  issn = {1557-9603},
  doi = {10.1109/TAES.2016.140952},
  urldate = {2026-07-02}
}

@Article{virtanen2020scipy,
  author     = {Virtanen, Pauli and Gommers, Ralf and Oliphant, Travis E. and Haberland, Matt and Reddy, Tyler and Co
urnapeau, David and Burovski, Evgeni and Peterson, Pearu and Weckesser, Warren and Bright, Jonathan and van der Walt,
 St{\'e}fan J. and Brett, Matthew and Wilson, Joshua and Millman, K. Jarrod and Mayorov, Nikolay and Nelson, Andrew R
. J. and Jones, Eric and Kern, Robert and Larson, Eric and Carey, C. J. and Polat, {\. I}lhan and Feng, Yu and Moore,
 Eric W. and VanderPlas, Jake and Laxalde, Denis and Perktold, Josef and Cimrman, Robert and Henriksen, Ian and Quint
ero, E. A. and Harris, Charles R. and Archibald, Anne M. and Ribeiro, Ant{\^o}nio H. and Pedregosa, Fabian and van Mu
lbregt, Paul},
  title      = {{SciPy} 1.0: fundamental algorithms for scientific computing in {Python}},
  doi        = {10.1038/s41592-019-0686-2},
  number     = {3},
  pages      = {261--272},
  urldate    = {2023-08-23},
  volume     = {17},
  copyright  = {2020 The Author(s)},
  journal    = {Nature Methods},
  month      = mar,
  shorttitle = {{SciPy} 1.0},
  year       = {2020},
}

@article{gopalsamyDiscoveryBenzisoxazolesPotent2008a,
  title = {Discovery of {{Benzisoxazoles}} as {{Potent Inhibitors}} of {{Chaperone Heat Shock Protein}} 90},
  author = {Gopalsamy, Ariamala and Shi, Mengxiao and Golas, Jennifer and Vogan, Erik and Jacob, Jaison and Johnson, Mark and Lee, Frederick and Nilakantan, Ramaswamy and Petersen, Roseann and Svenson, Kristin and Chopra, Rajiv and Tam, May S. and Wen, Yingxia and Ellingboe, John and Arndt, Kim and Boschelli, Frank},
  year = 2008,
  month = feb,
  journal = {Journal of Medicinal Chemistry},
  volume = {51},
  number = {3},
  pages = {373--375},
  publisher = {American Chemical Society},
  issn = {0022-2623},
  doi = {10.1021/jm701385c},
  urldate = {2026-07-23}
}
\end{document}